\documentclass[fleqn,10pt]{wlscirep}
\usepackage[utf8]{inputenc}
\usepackage[T1]{fontenc}
\usepackage{makecell}
\usepackage{multirow}
\usepackage{comment}
\title{Evidence of Inter-state Coordination amongst State-backed Information Operations}

\author[1,*]{Xinyu Wang}
\author[1]{Jiayi Li}
\author[1]{Eesha Srivatsavaya}
\author[1,*]{Sarah Rajtmajer}
\affil[1]{College of Information Sciences and Technology, The Pennsylvania State University, University Park, PA, USA}

\affil[*]{xzw5184@psu.edu, smr48@psu.edu}

\begin{abstract}
Since 2018, Twitter has steadily released into the public domain content discovered on the platform and believed to be associated with information operations originating from more than a dozen state-backed organizations. 
Leveraging this dataset, we explore inter-state coordination amongst state-backed information operations and find evidence of intentional, strategic interaction amongst thirteen different states, separate and distinct from within-state operations. We find that coordinated, inter-state information operations attract greater engagement than baseline information operations and appear to come online in service to specific aims. We explore these ideas in depth through two case studies on the coordination between Cuba and Venezuela, and between Russia and Iran.
\end{abstract}
\begin{document}

\flushbottom
\maketitle

\thispagestyle{empty}

\section*{Introduction}

The current reach of social media platforms and their affordances for cheap and easy content dissemination, profiling, and targeting, have established social media as a primary avenue for information operations -- efforts to manipulate public opinion by intentionally altering the information environment \cite{starbird2019disinformation}. A substantial literature has emerged studying the tactics and strategies of information operations, particularly on Twitter, where data has been widely available \cite{pierri2020investigating,soares2021hashtag,xia2019disinformation}. These studies have focused on campaigns attributed to individual states or state-backed organizations. To the best of our knowledge, no prior work has looked at collaboration amongst states in these efforts. Yet, examples of international collaboration for the dissemination of propaganda date back to the first and second World Wars \cite{casey1944goebbels,berelson1947detecting,waddington2007anti,cabrera2020international}. Our research explores evidence of inter-state coordination amongst state-backed information campaigns, operationalized through the following two research questions. 

\noindent \textbf{RQ1}: Do state-backed information campaigns operating on Twitter collaborate across states? If so, what distinguishes these efforts from internal information operations in terms of design, deployment, and impact?

\noindent \textbf{RQ2}: Can we categorize strategic and tactical mechanisms underlying inter-state information operations? E.g., specific roles of individual accounts in support of collusion?

We extract interaction networks amongst thirteen state-backed campaigns operating on Twitter between 2011 and 2021, perform static and dynamic pairwise analyses of observed activity, and highlight the varied structure of inter-state information operations through two case studies. Our findings indicate that inter-state operations attract greater engagement than intra-state operations. We suggest that the strategies employed by state-backed information operations serve to create and maintain a desirable information habitat, e.g., by engaging in ambient affiliation through common hashtags \cite{zappavigna2014enacting}, initiating network expansion for increased exposure \cite{zhang2021assembling}, and referencing controversial topics to gain attention \cite{garcia2018share}. Our findings represent the first insights into tactics and strategies underlying global cooperation and collusion amongst states in strategic information operations deployed through social media.

\section*{Related Work}
\subsection*{Strategic information operations} Starbird et al. \cite{starbird2019disinformation} use the term \emph{strategic information operations} (IO) to refer to efforts by individuals and groups, including state and non-state actors, to manipulate public opinion and change how people perceive events in the world by intentionally altering the information environment. Tracing their roots to TV and radio propaganda in the 20th century, and in some variation even much earlier \cite{posetti2018short}, the modern digital age and in particular today's social media landscape have enabled these efforts with unparalleled efficiency and have raised a unique set of questions around response \cite{rajtmajer2020automated}.  

A primary subset of information operations aims to disseminate inaccurate, incorrect, or misleading information, so-called \emph{disinformation}. Disinformation operations have been strongly associated with political campaigns \cite{niblock2022understanding}, focused primarily on social media manipulation of public opinion through bot accounts and paid workers \cite{bradshaw2019global,bradshaw2018challenging,arnaudo2017computational,ong2018architects}. In attempt to combat disinformation, scholars have focused efforts on developing technical solutions for automated detection of multimodal disinformation, e.g., fact-checking claims in text, identifying falsified images, and detecting deception through speech and facial expression in video \cite{vo2018rise,sathe2020automated,volkova2019explaining,garimella2020images,krishnamurthy2018deep}.

While the vast majority of studies reporting on information operations focus on individual state efforts, see e.g., \cite{dawson2019russia,myers2019china,merhi2021information}, or events, e.g., presidential elections \cite{ferrara2017disinformation,faris2017partisanship,grinberg2019fake}, a few have presented a global overview of information operations and therefore are most similar to our work here. Bradshaw and Howard's 2018 report provides an inventory of organized social media manipulation campaigns in 70 countries \cite{bradshaw2018challenging}. Niblock et al. have compiled comprehensive summary statistics and visualizations of all publicly-shared state-backed information operations on Twitter \cite{niblock2022understanding}. Our study likewise provides global analyses, but is unique in its focus on inter-state coordination.

\subsection*{Information operations as collaborative work}
Recent work in the Human-Computer Interaction (HCI) literature has highlighted the participatory nature of online information operations and interpreted the dissemination of manipulated information via the lens of online communities \cite{starbird2019disinformation}. Psychological theories, e.g., distributed cognition \cite{hutchins1995cognition}, offer theoretical bases for examining how social environments, such as mainstream social media platforms, impact collective behaviors among social ties and facilitate disinformation \cite{sarcevic2012beacons}. In particular, sociotechnical systems allow information operations to target, integrate with, and leverage the activities of online crowds - resulting in a combination of orchestrated and organic elements and behaviors. Schoch et al. \cite{schoch2022coordination} find that online political astroturfing consistently leaves similar patterns of coordination across distinct geographical settings and historical periods. The collaboration amongst state-backed accounts we describe in this work appears to be orchestrated. However, the audience with whom they engage and acts of implicit coordination with native users, e.g., via hashtagging, embedded URLs, is critical to our work.

\subsection*{Idea habitats}
Many mechanisms have been proposed to model the diffusion of misinformation and to expose environmental factors that drive false beliefs \cite{del2016spreading,wang2020fake}. Current studies capture a number of psychological, demographic, and environmental variables contributing to the acquisition and spread of misinformation \cite{wang2020fake,chen2015deterring,pennycook2020fighting,pennycook2021psychology}.
Work in cognitive science has suggested that the ways and extent to which individuals recall and distribute information depends on collections of environmental cues, so-called idea habitats\cite{berger2005idea}. These cues may include social and political context, linguistic characteristics, and topics of conversation. Common practices such as audience segmentation and micro-targeting represent efforts to nurture habitats receptive to particular narratives. Suitable habitats support self-reinforcing information flows, independent of content validity \cite{penney2017citizen}. In fact, studies have shown that false news spreads further, faster, and more broadly than legitimate news, a phenomenon which has magnified challenges to combat mis- and disinformation in recent years \cite{jakubowski2019s}. 
Our work dovetails with fundamental notions of idea habitat. We examine how accounts involved in information operations cooperate to establish context conducive to the distribution of misleading information.

\subsection*{Role analysis in social networks}
Prior work has analysed the different roles of actors/nodes in social network graphs using both graph structure-based and content-based approaches \cite{lee2014discovering}. Structural role analysis is often used for influence maximization tasks\cite{han2014balanced,lorrain1971structural} while content-based role analyses have found use in modeling the growth of online communities \cite{forestier2012roles,lee2014discovering,rundin2021multifaceted}. 
Our work uses a hybrid approach to define roles in inter-state information operations integrating network metrics and content-based analyses.

\section*{Dataset}
Since October 2018, Twitter has made public the tweets, media, and account-related information of users presumed to be involved in state-linked information operations, provided through the Twitter Moderation Research Consortium (TMRC). The TMRC suggests these users are engaged in \emph{manipulation that can be reliably attributed to a government or state-linked actor}\cite{TwitterIO}. We aggregate all account activity shared by the TMRC between 2007 and 2021. In total, this represents 23 state-backed information operations consisting of the full activity of 84,262 distinct accounts and approximately 120 million archived tweets.

During preprocessing, we made the following modifications to the complete dataset. We combined accounts designated by Twitter as linked to Egypt and UAE. Twitter's documentation subsequent to the release of these accounts indicated that much of their activity was attributed to an operation managed out of both countries targeting Qatar and Iran with messaging supportive of the Saudi government \cite{TwitterBlog}.  We did not include data from one release in March 2020 which Twitter attributed to ``Egypt, UAE and Saudia Arabia'' because attribution to a single country was not possible. This omitted subset of the data was relatively small (5350 accounts, 6.3\% of the total dataset). 
We did not find evidence of inter-state activity in content originating from Armenia, Bangladesh, Thailand, Tanzania, Mexico, Catalonia, Ghana, Nigeria, or Spain. These nine countries are therefore included in our dataset and analyses but not represented in the results, which focus on inter-state coordination. In sum, the number of tweets represented by these nine countries accounts for less than $0.2\%$ of the data. Dataset statistics are further detailed in the analyses below.

\section*{RQ1: Inter-state activity}
We use the terms ``coordination'' and ``coordinated operations'' to characterize purposeful collaboration in service to shared objectives. Informed by explanations of the dataset provided by the TMRC upon each data release, our analyses make the following assumptions: 1. All activities associated with accounts tagged by Twitter as participating in information operations are part of those operations; 2. All pairwise interactions between state-backed influence operation actors are coordinated information operations/platform manipulation. 
Evidence of inter-state coordination is informed by static and dynamic network and content analyses across state-linked accounts.

\subsection*{Inter-state interaction network}
\begin{figure}[ht]
    \centering
    \includegraphics[width=0.8\linewidth]{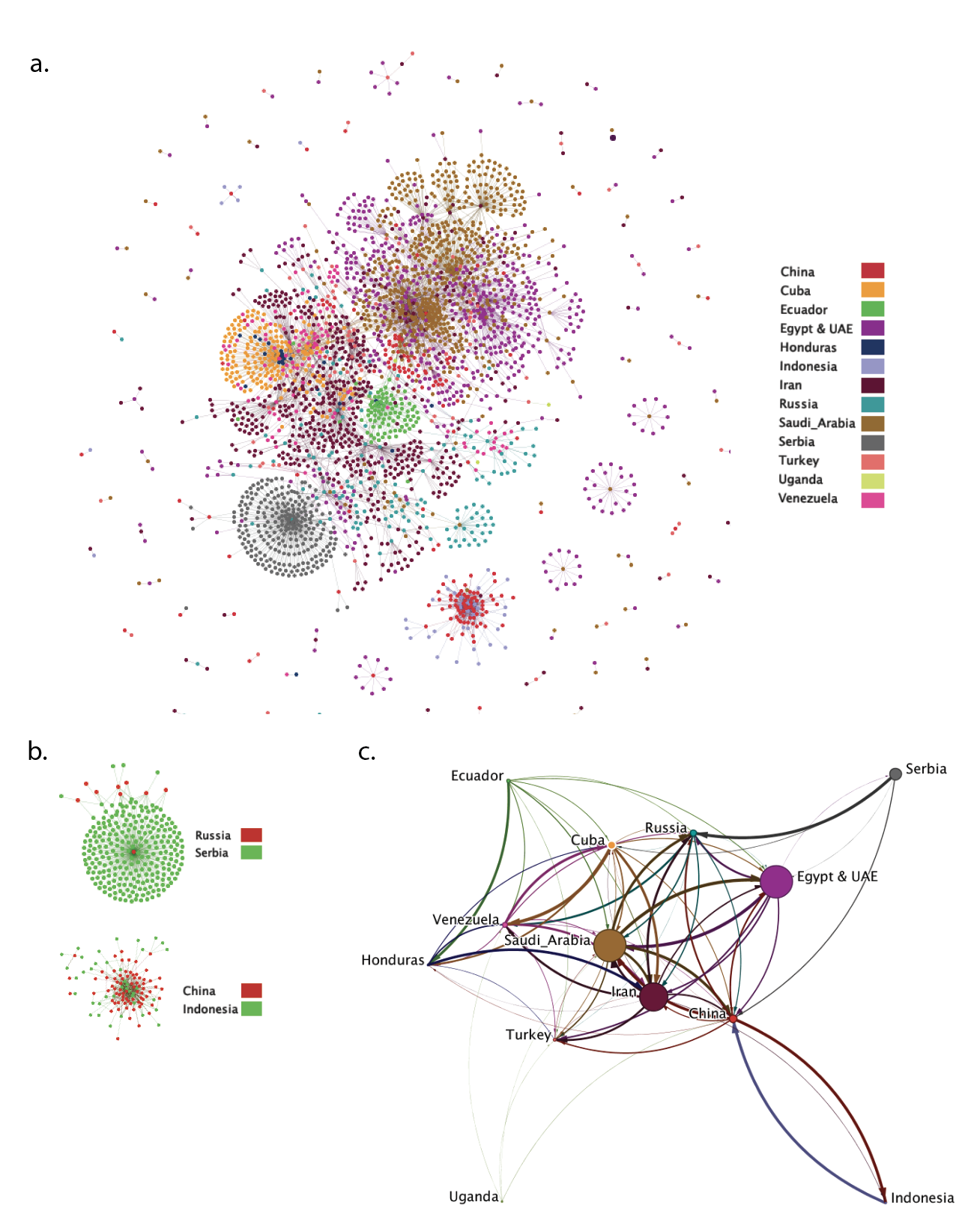}
    \caption{a. Global inter-state interaction network; b. Observed network motifs; c. Inter-state interactions, aggregated by country}
    \label{fig:network}
\end{figure}

We build a global inter-state interaction network amongst state-linked accounts (Figure~\ref{fig:network}a). Nodes represent accounts and directed edges represent retweets, replies, mentions, and quotes, between 2011 and 2021. Node color corresponds to country and edge color matches source node. We observe two predominant substructures within the network. The first is a radiating pattern, consisting of one or a few central nodes with high out-degree centrality (e.g., Figure~\ref{fig:network}b(i)). This motif appears for countries with either dominating in-degree centralization or out-degree centralization. Central nodes function as either content creators or self-promoters surrounded by a substantial number of followers and amplifiers to disseminate content and establish new social ties. The contrasting motif is balanced with similar in- and out-degree (e.g., Figure~\ref{fig:network}b(ii)). Figure~\ref{fig:network}(c) distills the inter-state interaction network via aggregation by country. Node size is proportional to log-scaled number of associated accounts. Edge width is proportional to log-scaled number of interactions in each pair-wise coordination and edge color matches source node. We note that Cuba, Serbia, and Ecuador appear to use retweets and replies to connect with other states for network expansion and content promotion. Whereas, Venezuela, Russia, Turkey, and Iran are predominantly the target of interactions. Accounts linked to these countries disseminate relatively more original content. Indonesia, Egypt \& UAE, China, and Saudi Arabia exhibit more balanced structures. We calculate the reciprocity of each state in the weighted network, decomposing dyadic fluxes into a fully reciprocated component and a fully non-reciprocated component \cite{squartini2013reciprocity}. Reciprocity levels are 0.944, 0.924, 0.844, and 0.777 for Indonesia, Egypt \& UAE, China, and Saudi Arabia, respectively. With the exception of Honduras, where we observe a comparable number of interactions as source and target, 94.9\% of global outgoing interactions are directed to Russia, and 88.3\% of global in-coming interactions originate from Ecuador (reciprocity: 0.0079).
\begin{table}[ht]
\small
    \centering
    \begin{tabular}{|c|c|c|c|c|c|c|c|}
        \hline
         \textbf{Country} & \textbf{\makecell{Inter-state\\actors}}&\textbf{\makecell{Inter-state\\interactions\\( source)}}&\textbf{\makecell{Inter-state\\interactions\\(target)}} & \textbf{\makecell{Intra-state\\interactions}}&\textbf{\makecell{External\\interactions \&\\isolates}}&\textbf{\makecell{Total \\actors}}&\textbf{\makecell{Total \\tweets}}\\
         \hline
           Cuba(CU)&181&6,649 &1,027&742,654&4,055,338&526&4,805,668 \\
         \hline
         Venezuela(VE)&114&1,521&6,131&219,973&10,368,229&2,261&10,595,854\\
           \hline
           Iran(IR)&569&1,528&3,275&905,384&9,524,551&7,025&10,434,738\\
           \hline
           Russia(RU)&143&356&5,227&352,943&4,124,305&1,741&4,482,831\\
           \hline
           Serbia(RS)&254&3,267 &1&5,750,528&7,968,305 &8,558&13,722,101\\
           \hline
           Turkey(TR)&72&5&359& 2,600,704& 12,750,040&7,340 &15,351,108\\
           \hline
           Indonesia(ID)&40&2,683&2,539 & 28,439 &2,669,174&795& 2,702,835\\
           \hline
           Honduras(HN)&23&1,064 &1,059& 37,529&1,126,426&3,104& 1,166,078\\
           \hline
           Ecuador(EC)&87&973 &0&23,456 &675,811&1,019& 700,240\\
           \hline
           China(CN)&176&4,251 &4,691&294,616&13,964,665&31,119& 14,268,223\\
           \hline
           Saudi Arabia(SA)& 649&8,650&6,813&225,499&32,047,515&5,968& 32,288,477\\
           \hline
           Egypt \& UAE(EU)&570& 5,187& 5,017&706,047 &8,764,523 &7,060& 9,480,774\\
           \hline
           Uganda(UG)&3&5 &0&62,857&461,219&418& 524,081\\

         \hline
    \end{tabular}
    \caption{Count of inter- and intra-state interactions by country. Inter-state: source and target nodes associated with \emph{different} state-linked operations; Intra-state: source and target nodes associated with the same state-linked operation; External \& isolates: target node is not identified by Twitter as a state-linked actor, or content does not retweet/mention other actors. \emph{Note: External interactions and isolates do not inform the analyses provided in this work; the count is provided for context.}
    }
    \label{tab:interactions}
\end{table}

\subsection*{Temporal analysis of inter- and intra-state activity}

\begin{figure}[ht]
    \centering
    \includegraphics[width=\linewidth]{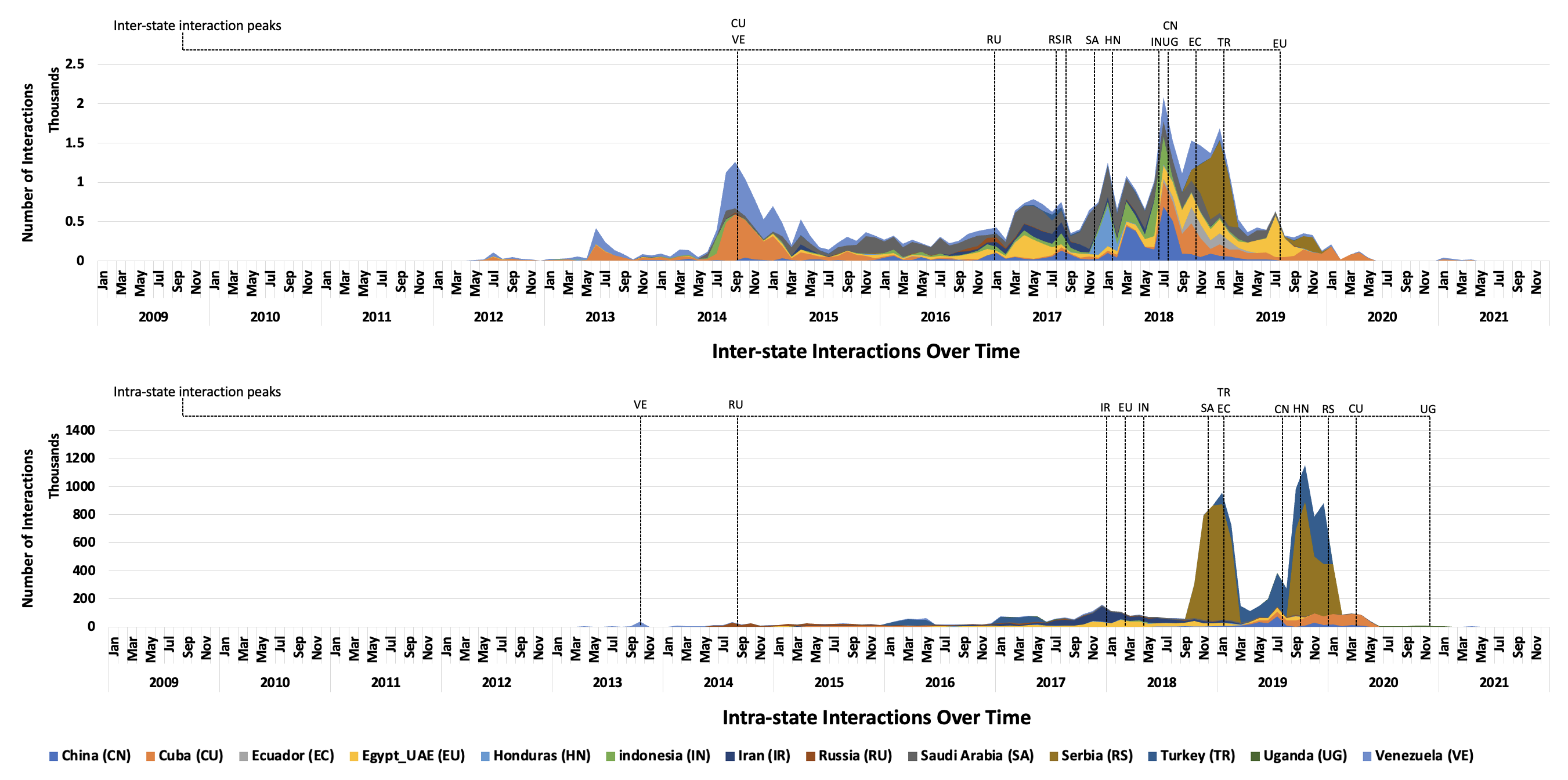}
    \caption{Cumulative intra- and inter-state interactions, over time.}
    \label{fig:total_dyn}
\end{figure}

Figure~\ref{fig:total_dyn} shows the cumulative inter- and intra-state interaction counts over time, along with labeled interaction peaks for each of the 13 states. We observe that initial inter-state activity lags behind intra-state activity (avg. lag approximately 2 years). Notably, a majority of inter-state interactions reach peak activity synchronously in late 2017 and 2018. Comparing inter- and intra-state interactions for each state individually, we find that 10 out of the 13 countries in our dataset have substantially different temporal patterns. That is, in most cases, inter-state operations do not occur concurrently with intra-state operations. Rather, they represent what appears to be a separate strategic operation. Seven peaks in inter-state and intra-state activities occur more than one year apart, three occur three to twelve months apart, two peaks occur within three months of one another, and one occurs simultaneously within the same month).

\subsection*{Measuring engagement with inter-state activity}
Observing that one aim of inter-state coordination appears to be increased visibility, we measure differences in engagement statistics between inter- and intra-state activity using a two sample T-test. We perform \emph{a priori} power analysis to determine the minimum sample size, set the significance level $\alpha$ to $0.05$, power to $0.8$, and Cohen's d effect size to $0.2$. We obtain a minimum effective sample size of $394$ to perform target statistical testing. Using this threshold value, we filter out two states (Turkey and Uganda) with less than $394$ inter-state interactions. Given the significantly smaller fraction of quotes in the dataset (the number of quote tweets in the dataset is less than the effective sample size), we select likes, retweets, and replies as three indices for comparative study. We perform a Welch's T test to determine if observed differences are significant. Table~\ref{tab:engagement} gives these results. 
In 7 of 11 states, inter-state interactions receive more likes on average than intra-state interactions, 5 of which are significant. Similar trends hold for retweets and replies. We observe that countries like Venezuela, China, Indonesia, and Egypt and UAE which are more active in inter-state coordination also apply tactics discussed later (RQ2).
\begin{table}[ht]
\small
    \centering
    \begin{tabular}{|c|c|c|c|c|c|c|c|c|c|c|c|c|}
        \hline
         \textbf{Country} & \textbf{\makecell{Avg\\ like}} &\textbf{\makecell{p}}&\textbf{\makecell{T}}&\textbf{\makecell{d}}&\textbf{\makecell{Avg\\ retweet}}&\textbf{\makecell{p}}&\textbf{\makecell{T}}&\textbf{\makecell{d}}&\textbf{\makecell{Avg\\ reply}}&\textbf{\makecell{p}}&\textbf{\makecell{T}}&\textbf{\makecell{d}}\\
         \hline
           VE(Inter)&0.401&\multirow{2}{*}{\textbf{0.000}}&\multirow{2}{*}{9.218}&\multirow{2}{*}{0.566**}&0.440&\multirow{2}{*}{\textbf{0.000}}&\multirow{2}{*}{4.737}&\multirow{2}{*}{0.096}&0.128&\multirow{2}{*}{\textbf{0.000}}&\multirow{2}{*}{6.038}&\multirow{2}{*}{0.478*} \\
           \cline{1-2}\cline{6-6}\cline{10-10}
           VE(Intra)&0.039&&&&0.168&&&&0.015&&&\\
           \hline
           \hline
           CU(Inter)&0.539&\multirow{2}{*}{0.169}&\multirow{2}{*}{1.375}&\multirow{2}{*}{0.057}&0.366&\multirow{2}{*}{0.794}&\multirow{2}{*}{0.261}&\multirow{2}{*}{0.011}&0.125&\multirow{2}{*}{\textbf{0.000}}&\multirow{2}{*}{8.420}&\multirow{2}{*}{0.314*}\\
           \cline{1-2}\cline{6-6}\cline{10-10}
           CU(Intra)&0.351&&&&0.332&&&&0.020&&&\\
           \hline
           \hline
           CN(Inter)
           &0.035 &\multirow{2}{*}{\textbf{0.000}}&\multirow{2}{*}{4.438}&\multirow{2}{*}{0.063}&0.058&\multirow{2}{*}{\textbf{0.000}}&\multirow{2}{*}{6.608}&\multirow{2}{*}{0.027}&0.351&\multirow{2}{*}{\textbf{0.000}}&\multirow{2}{*}{31.720}&\multirow{2}{*}{0.902***}\\
           \cline{1-2}\cline{6-6}\cline{10-10}
           CN(Intra)&0.016&&&&0.016&&&&0.025&&&\\
           \hline
           \hline
           IR(Inter) &0.056&\multirow{2}{*}{\textbf{0.000}}&\multirow{2}{*}{-6.142}&\multirow{2}{*}{-0.034}&0.052&\multirow{2}{*}{0.668}&\multirow{2}{*}{0.429}&\multirow{2}{*}{0.007}&0.101&\multirow{2}{*}{\textbf{0.000}}&\multirow{2}{*}{6.937}&\multirow{2}{*}{0.304*}\\
           \cline{1-2}\cline{6-6}\cline{10-10}
           IR(Intra)&0.159&&&&0.044&&&&0.018&&& \\
            \hline
           \hline
           RS(Inter)
           &0.000 &\multirow{2}{*}{\textbf{0.000}}&\multirow{2}{*}{-125.492}&\multirow{2}{*}{-0.053}&0.000&\multirow{2}{*}{\textbf{0.000}}&\multirow{2}{*}{-97.538}&\multirow{2}{*}{-0.041}&0.000&\multirow{2}{*}{\textbf{0.000}}&\multirow{2}{*}{-192.692}&\multirow{2}{*}{-0.082}\\
          \cline{1-2}\cline{6-6}\cline{10-10}
          RS(Intra)&0.012&&&&0.008&&&&0.012&&&\\
           \hline
           \hline
           RU(Inter)
           &0.007 &\multirow{2}{*}{\textbf{0.000}}&\multirow{2}{*}{-4.570}&\multirow{2}{*}{-0.010}&0.000&\multirow{2}{*}{\textbf{0.000}}&\multirow{2}{*}{-10.009}&\multirow{2}{*}{-0.017}&0.018&\multirow{2}{*}{\textbf{0.000}}&\multirow{2}{*}{-1.938}&\multirow{2}{*}{-0.084}\\
          \cline{1-2}\cline{6-6}\cline{10-10}
          RU(Intra)&0.034&&&&0.021&&&&0.038&&&\\
        \hline
           \hline
           ID(Inter)
           & 0.034&\multirow{2}{*}{\textbf{0.003}}&\multirow{2}{*}{2.981}&\multirow{2}{*}{0.040}&0.072&\multirow{2}{*}{\textbf{0.000}}&\multirow{2}{*}{6.476}&\multirow{2}{*}{0.111}&0.559&\multirow{2}{*}{\textbf{0.000}}&\multirow{2}{*}{33.782}&\multirow{2}{*}{1.425***}\\
          \cline{1-2}\cline{6-6}\cline{10-10}
          ID(Intra)&0.020&&&&0.019&&&&0.067&& &\\
          \hline
           \hline
           SA(Inter)
           &0.037&\multirow{2}{*}{\textbf{0.039}}&\multirow{2}{*}{2.064}&\multirow{2}{*}{0.012}&0.068&\multirow{2}{*}{0.138}&\multirow{2}{*}{1.482}&\multirow{2}{*}{0.007}&0.034&\multirow{2}{*}{\textbf{0.000}}&\multirow{2}{*}{-4.935}&\multirow{2}{*}{-0.027}\\
          \cline{1-2}\cline{6-6}\cline{10-10}
          SA(Intra)&0.026&&&&0.039&&&&0.050&&&\\
          \hline
           \hline
           EC(Inter)
           & 0.020&\multirow{2}{*}{0.995}&\multirow{2}{*}{-0.007}&\multirow{2}{*}{0.000}&0.002&\multirow{2}{*}{0.040}&\multirow{2}{*}{-2.059}&\multirow{2}{*}{-0.035}&0.000&\multirow{2}{*}{\textbf{0.000}}&\multirow{2}{*}{-6.803}&\multirow{2}{*}{-0.046}\\
           \cline{1-2}\cline{6-6}\cline{10-10}
          EC(Intra)&0.020&&&&0.007&&&&0.018&&&\\
          \hline
           \hline
           HN(Inter)
           &0.450 &\multirow{2}{*}{0.611}&\multirow{2}{*}{0.509}&\multirow{2}{*}{0.004}&0.619&\multirow{2}{*}{\textbf{0.000}}&\multirow{2}{*}{5.879}&\multirow{2}{*}{0.148}&0.030&\multirow{2}{*}{0.212}&\multirow{2}{*}{-1.250}&\multirow{2}{*}{-0.016}\\
           \cline{1-2}\cline{6-6}\cline{10-10}
          HN(Intra)&0.416&&&&0.182&&&&0.039&&&\\
          \hline
           \hline
           EU(Inter)
           &0.073&\multirow{2}{*}{\textbf{0.000}}&\multirow{2}{*}{6.683}&\multirow{2}{*}{0.100}&0.076&\multirow{2}{*}{\textbf{0.000}}&\multirow{2}{*}{5.774}&\multirow{2}{*}{0.114}&0.036&\multirow{2}{*}{\textbf{0.000}}&\multirow{2}{*}{8.463}&\multirow{2}{*}{0.148}\\
           \cline{1-2}\cline{6-6}\cline{10-10}
          EU(Intra)&0.012&&&&0.008&&&&0.012&&&

          \\

         \hline
    \end{tabular}
    \caption{Comparison of inter- and intra-state engagement. p: p-value; T: T-score; d: Cohen's d value; *: small effect size, **: medium effect size, ***: large effect size; p-values in bold indicate significance after FDR correction. 
    }
    \label{tab:engagement}
\end{table}

We additionally perform engagement comparisons between inter-state interactions and state-backed accounts' interactions with external accounts, i.e., accounts not tagged by Twitter as state-backed actors, presumed ``normal'' accounts). 
As the data is imbalanced (see Table \ref{tab:interactions}), we randomly sample external interactions matching the observed number of inter-state interactions. With a similar level of variance, we perform a regular two-sample T-test. Results are provided in Table S3 of Supplementary Material. We observe 
a general pattern of more likes and retweets associated with external accounts. This is expected as external accounts have greater visibility, e.g., news outlets, and likes and retweet counts are derived from the original post. However, number of replies associated with inter-state interactions is substantially \emph{greater} than those associated with external interactions, indicating success of inter-state coordination to prompt meaningful engagement (e.g., Cuba, China, and Indonesia).

\section*{RQ2: Strategic and tactical mechanisms}
We study the strategic use of network structure and shared content in service to inter-state coordination. These are explored in detail through two case studies -- coordination between Cuba and Venezuela and between Russia and Iran. These examples are selected to highlight the diversity of structural and functional activity we observe across the dataset. In the case of Cuba and Venezuela, we observe relatively bi-directional interaction; both countries serve as source and target of coordinated activity. We also observe administrators playing distinct roles in the campaign. Russia and Iran's coordinated operations, on the other hand, are at a larger scale and structurally very different.

\subsection*{Ambient Affiliation}
Implicit association among social network actors is facilitated through hashtagging, a phenomenon which has been studied in the sociolinguistics literature as \emph{ambient affiliation} \cite{zappavigna2014enacting}. These indirect interactions enhance visibility of users' discourse through search \cite{zappavigna2011ambient}. The social role of hashtagging is to facilitate the establishment of ad hoc social interaction groupings or subcommunities, which constitute a temporal habitat for information operations. Hashtagging has been employed and proven effective across platforms, from ``influencers'' and organizations to disinformation operations, for acquiring followers and increasing exposure \cite{erz2018hashtags,conway2013twitter,martin2016hashtags,starbird2019disinformation}.
We suggest that inter-state ambient affiliation is used by information operations to create idea habitats conducive to information spread. 

\begin{figure}[ht]
    \centering
    \includegraphics[width=\linewidth]{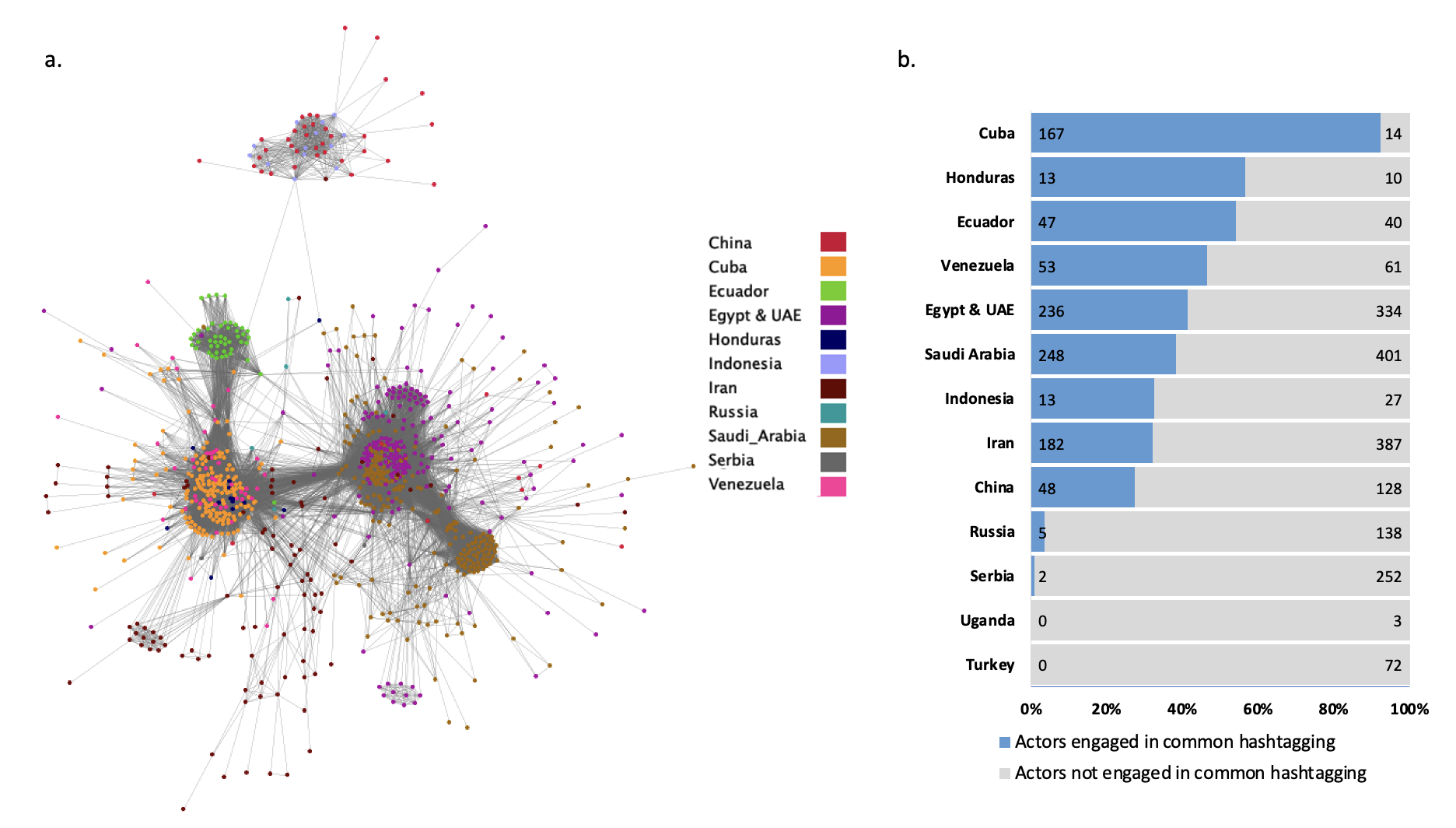}
    \caption{ a. Inter-state hashtag network; \emph{Note: Components of size $\leq 20$ are suppressed for clarity.} 
 b. Actor engagement within inter-state hashtag network, by country.}
    \label{fig:hashtag}
\end{figure}

We construct the inter-state hashtag network (Figure~\ref{fig:hashtag}(a)). Nodes in the network represent accounts that both engage in hashtagging and are involved in inter-state coordination; Undirected edges indicate use of common hashtags by these accounts. 
There are $1,014$ nodes and $61,010$ edges in the network (density = $0.594$; diameter = $11$). 
The largest connect component contains $861$ nodes and $60,510$ edges (density = $0.0817$).  
Use of ambient affiliation is variable across operations (Figure~\ref{fig:hashtag}(b)). While pervasive within Cuban operations, Russia, Serbia, and Turkey use this tactic only negligibly despite having large-scale operations. Notably, engagement in ambient affiliation appears correlated with greater engagement (see Table ~\ref{tab:engagement}, e.g., we see greater engagement with content from Cuba and Honduras than from Russia and Serbia.)

\begin{figure}[ht]
    \centering
    \includegraphics[width=0.4\linewidth]{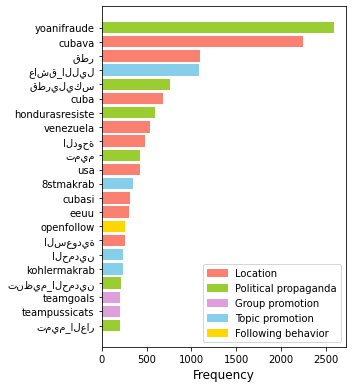}
    \caption{Top hashtags used in inter-state IO. (English translations, from top: Yoani fraud, Cuba goes, Qatar, love the night,
Qatarileaks, Cuba, Honduras resist, Venezuela, Doha, Tamim, USA, 8stmakrab*, Cuba yes, USA, openfollow, Saudi Arabia, Al-Hamdan, Kohlermakrab*, organization\_Hamdeen, Team Goals, Team Pussicats, Tamim\_shame; \emph{*Note: Makrab is the abbreviation of the phrase "Malam Keakraban" in Indonesian which means friendship night.})}
    \label{fig:hashtag_freq}
\end{figure}

Globally, we identify $1,148$ unique hashtags that occur within inter-state activity a total of $33,119$ times.
Figure~\ref{fig:hashtag_freq} lists and categorized the hashtags which appear in at least $200$ inter-state interactions. We observe pervasive, intentional exploitation of political controversy within inter-state hashtagging behavior. In the case of several prominent collaborative operations, a majority of inter-state interactions target specific political events (e.g. Honduras and Iran: 2017 Honduras’ Election Crisis; Iran and Russia: 2016 U.S election; Iran and Venezuela: 2017 Venezuelan Protests). Other coordination activities incorporate media outlets associated mostly with unsubstantiated news. 

In addition, we observe the use of hashtagging for network expansion, e.g., \#syts, \#openfollow, \#siguemeytesigo (follow me and I'll follow you). This tactic appears to take one of two forms: (1) explicitly requesting followers; and (2) using mentions and tags for penetration into new communities. 


\subsection*{Taxonomizing roles within inter-state operations}
Within inter-state operations, we observe that different users/accounts appear to have different patterns of behavior. We contextualize these differences through the lens of role analysis, defining primary roles as follows: 
\begin{itemize}
 \item \textbf{Administrator.} Manages operations of the information campaign. Administrators self-identify as group leaders through profile information and shared content. 
 \item \textbf{Influencer.} A hub of the operation with high in-degree centrality. Typically, an influencer is the source of the information who exploits fake news sites and may have multiple similar accounts in the network to avoid takedown. We define users with in-degree greater than 10 as the influencers in the network.
    \item \textbf{Promoter.} Primarily promotes content for enhanced visibility and engagement. We define users with out-degree greater than 10 as promoters in the network.
    \item \textbf{Broker.} A gatekeeper, connecting multiple communities/organizations with relatively high in-degree and out-degree centrality.
    We identify users who meet criteria for both promoter and influencer as brokers. 
    \item \textbf{Follower.} An actor with minor (observable) impact within the operation. Users that are not identified within aforementioned roles are categorized as followers.
\end{itemize}

\noindent We leverage this taxonomy in the case studies which follow. The two case studies are selected for their diversity with respect to structure and content.

\subsection*{Case Study 1: Coordination between state-linked accounts from Cuba and Venezuela}

\begin{figure}[ht]
    \centering
    \includegraphics[width=0.95\linewidth]{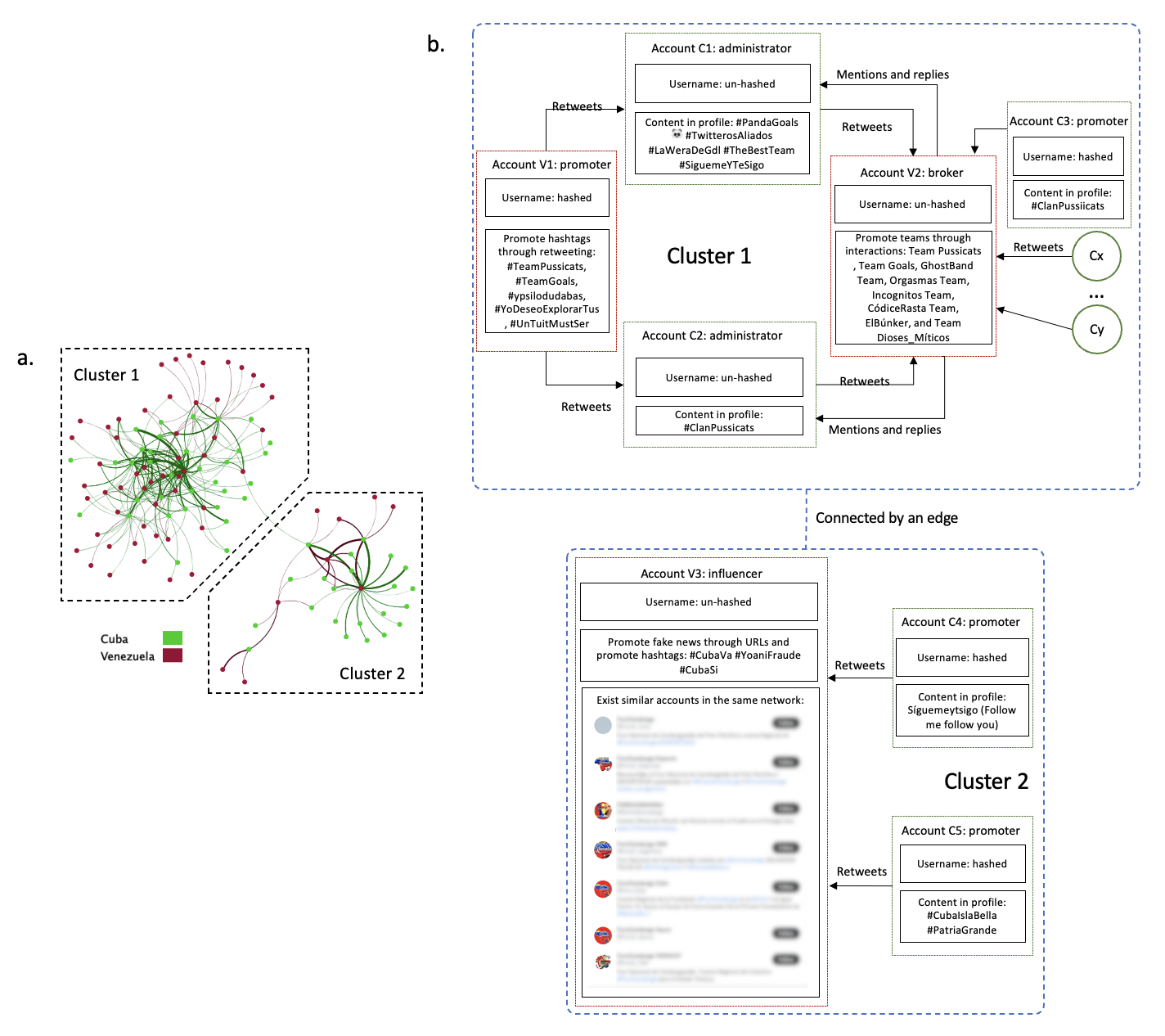}
    \caption{a. Inter-state interaction network between Cuba and Venezuela; b. Schematic structure of inter-state disinformation operations between Cuba and Venezuela (usernames are hashed for accounts with fewer than $5,000$ followers). }
    \label{fig:CUVE}
\end{figure}

\subsubsection*{Network structure}
We construct the inter-state interaction network between Cuban and Venezuelan state-linked accounts. We observe $6,469$ interactions between $56$ Cuban and $62$ Venezuelan accounts. Notably, the network has two well-connected clusters connected by a single edge (see Figure~\ref{fig:CUVE}(a). We observe a relatively balanced network structure between Cuba and Venezuela where the interactive pattern is bidirectional ($5,464/6,649$ of Cuba's out-degree interactions point to Venezuela, and $1,005/1,027$ Venezuela's out-degree interactions point to Cuba).  In each cluster, a small subset of Cuban and Venezuelan nodes dominate activity while remaining nodes connect with them through retweets and mentions. This network structure is a trademark of Cuba's inter-state operations. We observe that Cuba's intra-state interaction network, by contrast, has greater connectivity and more uniform in-/out-degree sequences.

\subsubsection*{Content analysis}

Notably, Cuban and Venezuelan information operations appear to center around structured \emph{teams}, and each team has a self-identified administrator. 

\noindent\emph{Cluster 1. }
In Figure~\ref{fig:CUVE}(b), we dive deeply into the inter-state operation structure. Within Cluster 1, we identify five representative nodes, two attributed to the Venezuelan campaign and three from Cuba. Representative nodes are selected as those with the most within-cluster interactions. The two team leaders are also identified though their user profiles and account descriptions. V1 is a Venezuelan node with high out-degree centrality, and the two Cuban nodes with which V1 frequently interacts are designated C1 and C2. 

Actors C1 and C2 are administrators in TeamGoal and TeamPussicats, respectively. The majority of unilateral interactions from V1 to C1 and C2 take the form of direct retweets and retweets from others that mention C1 or C2. Manual content analyses suggest that the primary objective of these two Cuban actors is to promote their team and its members through establishing unique sets of emojis and hashtags that symbolize their team identities and consistent mentioning of the team leaders.
By retweeting C1 and C2 as well as other members of their teams, V1 serves as a promoter of their content. Venezuelan node V2 has bilateral coordination with both C1 and C2. The majority of V2's activities are replies to users or tweets with direct mentions. These two sets of tweets focus mostly on promoting members of teams including C1 and C2 and others, e.g., GhostBand Team, Orgasmas Team, Incognitos Team, CódiceRasta Team, ElBúnker, and Team Dioses\_Míticos. V2 acts as a broker between the aforementioned groups, facilitating communication and collaboration between the teams and the team members.

\noindent\emph{Cluster 2. }
Within Cluster 2, we identify node V3 as an influencer. V3 primarily distributes fake news via the use of URLs and hashtags that receive considerable engagement ($65.26\%$ of total tweets from Cuba are connected to V3).
We note that numerous accounts that closely resemble this suspended account still exist in the current social network to avoid being taken down by Twitter, as shown in Cluster 2 of Figure~\ref{fig:CUVE}(b). V3 is linked to C4, C5, and several other Cuban promoter accounts, through both explicit (retweeting) and implicit interactions (common hashtags).

Role analysis of accounts within the inter-state Cuban-Venezuelan operations (see Supplementary Material) suggests that, broadly speaking, Venezuelan accounts act as influencers, while Cuban accounts primarily promote content shared by Venezuelan accounts.

\subsubsection*{URL analysis} We collect account profile information and tweets of all accounts engaged in inter-state operations between Cuba and Venezuela. In total, there are 16 accounts ($13.56\%$ of actors from both countries) whose profiles contain URLs. Further, we identify 1,913 unique URLs within their tweets, occurring over 2,553 interactions (39.47\% of total interactions) (see Supplementary Material). Invalid URLs that include broken links and cannot be manually identified by name are removed. Then, we manually verify the status of each link, categorizing each as active if the link is still functioning and inactive if the content has been removed or the account has been set to private. A substantial number of URLs are invalid (70.10\%), indicating that most were temporary. We observe that the majority of valid URLs in profiles redirect to accounts on other social networking sites like Facebook, Instagram, and YouTube (9.27\% of total valid URLs, 37.74\% of which are still active), and blogs with little or no regulations (28.85\% of total valid URLs, 100\% of which are still active). URLs within tweets often direct to politically-oriented news (57.00\% of total valid URLs), e.g., Telesurtv. Some also point to social media content management applications such as Twitlonger to bypass the character limit and Twitpic for picture archiving (still accessible after the account has been taken down), as well as to manage followers from social media, likely for open follow practices, e.g., Tuitil. 

\subsection*{Case Study 2: Coordination between state-linked accounts from Russia and Iran}

\subsubsection*{Dynamic network structure}
\begin{figure*}[ht]
    \centering
    \includegraphics[width=0.95\linewidth]{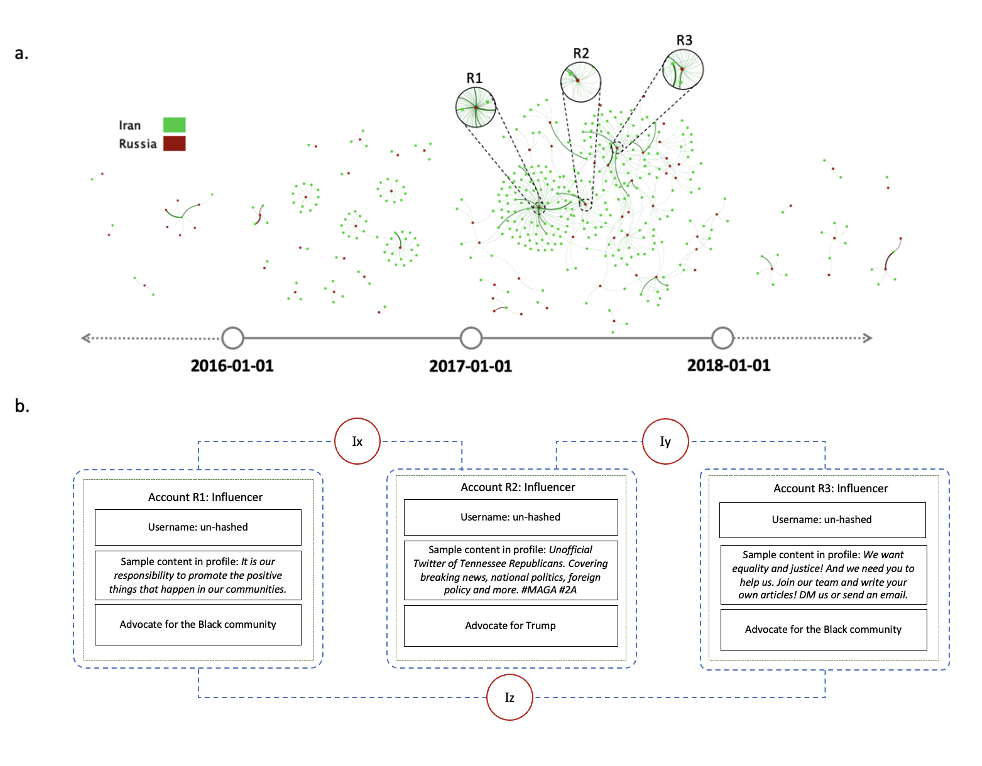}
    \caption{a. Dynamic inter-state interaction networks between Russia and Iran; b. Schematic structure of inter-state information operations between Russia and Iran.}
    \label{fig:RUIR}
\end{figure*}

We construct a dynamic view of  inter-state operations between Russia and Iran through four interaction networks, see Figure~\ref{fig:RUIR}(a). We aggregate all interactions prior to 2016 and after 2018, since the majority of interactions occur between 2016 and 2017; 2016 and 2017 are represented as snapshots. A total of 54 Russian accounts and 329 Iran accounts are represented in the network. We observe several clusters that contain one Russian node and multiple Iranian nodes, in a radiating pattern. The radiating network topology is well-suited for executing strategic aims such as news media distribution and social network expansion. 

Russian and Iranian accounts are dynamically involved in inter-state coordination globally (see Supplementary Material). These statistics are calculated over year-long network snapshots, beginning in 2011. We observe temporal uniformity for Iranian inter-state information operations; specifically, accounts reach out-degree interaction peak at around the same time (see Supplementary Material). This may suggest that inter-state operations were/are conducted concurrently. Although Iran has more incoming than outgoing interactions (see Table \ref{tab:interactions}), Iran also exhibits particularly many outgoing interactions with Russia. Upon closer inspection, we observe that the reason behind Iran's high in-degree centrality is that several central news outlets, e.g., Iranian state-controlled Hispantv, is frequently and consistently retweeted by accounts linked to Cuba, Venezuela, and Honduras (99.01\%, 99.58\%, and 99.7\% of the total tweets, respectively). Thus, Iran may be passively engaged in those interactions. However, in its coordination with Russia, Iranian actors actively disseminate content created by Russian actors, e.g., targeting the U.S. 2016 election.

\subsubsection*{Content analysis} 
Figure~\ref{fig:RUIR}(b) explores the primary structure of inter-state operations between Russia and Iran. We observe that Iranian accounts actively initiate interaction with Russian actors; approximately 96.81\% of activity in the inter-state interaction network are retweets of Russian accounts, the majority of which are affiliated with the Internet Research Agency (IRA). Manual analyses of tweets and profile descriptions indicate that content is primarily focused on political and politicized topics. This is in line with prior work examining Russian information operations targeting the 2016 U.S. election \cite{benkler2018network,berghel2017oh}. 
For instance, the profile description of the central node R2 is: 
\begin{quote}
Unofficial Twitter of Tennessee Republicans. Covering breaking news, national politics, foreign policy and more. \#MAGA \#2A
\end{quote}
In addition to U.S. election politics, inter-state activity also focuses on racial issues (e.g., R1 and R3). Iranian actors participate in these processes by disseminating material previously posted by Russian actors with high centrality , and connecting all three satellite communities.

As in Case Study 1, we identify account roles in the Russia-Iran inter-state campaign (see Supplementary Material). Russia and Iran differ significantly in terms of number of influencers and followers; Russian actors serve primarily as source content for the Iranian community.

\subsubsection*{URL analysis} 
Also as in Case Study 1, we collect profile information and tweets from all accounts engaged in inter-state operations between Russia and Iran. In total, there are just 7 accounts (1.83\% of actors from both countries) that contain URLs within their profile descriptions. We identify 35 unique URLs within shared tweets over 77 total interactions (15.28\% of total interactions). We categorize these URLs by type/platform and active status (see Supplementary Material). Among valid URLs, we see extensive reference to international political news from the United States, the United Kingdom, and Iraq (31.43\%). While the majority of linked social media accounts are now defunct (75\% for profile URLs and 100\% for content URLs) URLs to most news outlets remain active and complement tweet content (68.75\% for content URLs). 

\section*{Discussion and Conclusions}

Rapid and accurate detection of information operations remains a hard problem for social media platforms globally. Evidence that state-linked operations may collaborate or collude to improve the efficacy of their campaigns adds complexity to that challenge. Our work provides such evidence, and therefore informs ongoing efforts to detect and mitigate the impact of information operations deployed through social media. 
We have highlighted some recurring strategies and tactics employed by inter-state information operations on Twitter. We have observed that a substantial subset of coordinated inter-state activity can be identified as supportive of explicit aims, e.g., targeting high-stakes political events or seeking additional visibility. Regardless of motivation, it appears that inter-state activities are carried out separately from intra-state operations, resulting in a distinctive information ecosystem, or idea habitat. Relatedly, we discover that a majority of inter-state operations exploit ambient affiliation through hashtagging, and that individual accounts in the network may be tasked with distinct roles in some operations architectures. Overarchingly, our findings suggest that information operations represent collaborative work, not only at the individual level but also at the state level. Notably, our analyses also reveal that country size is not necessarily a determinant of the scale of observed inter-state activity. Smaller countries demonstrate the ability to engage in systematic coordination, strategically expanding their internal operations.

The scope of the current work is constrained in several ways. Our analyses make assumptions about the accuracy of identified accounts and about their activity, e.g., that all account activity serves ongoing operations. We assume that substantial observed interaction indicates strategic coordination. Ultimately, the emergence of inter-state coordination and the ways in which observed activities are moderated through planning remains an open question.  
Furthermore, the extent to which current insights can be leveraged for the advancement of automated approaches for detection of inter-state information operations will likely depend on the uniqueness of inter-state interaction patterns vs., e.g., standard content promotion strategies employed by traditional organizational accounts seeking visibility and influence. Future work would benefit from designing studies to compare these phenomena.

The inter-state activity we uncover here is likely a very small segment of much larger-scale, dynamic, cross-platform, multi-media, partially-observable coordinated operations.  Our findings raise many more questions regarding the offline interactions which underlie observed activity, the ways in which they are facilitated, and the broader political agenda which they serve. The answers to these questions will be distinct across countries and over time. Our work therefore highlights the critical role of policy and international relations in this space. It also suggests that whether and how states cooperate to respond to information operations on social media will require a transdisciplinary research and policy agenda bringing together computational and social scientists, policy makers, and stakeholders.

\bibliography{main.bib}

\begin{thebibliography}{10}
\urlstyle{rm}
\expandafter\ifx\csname url\endcsname\relax
  \def\url#1{\texttt{#1}}\fi
\expandafter\ifx\csname urlprefix\endcsname\relax\def\urlprefix{URL }\fi
\expandafter\ifx\csname doiprefix\endcsname\relax\def\doiprefix{DOI: }\fi
\providecommand{\bibinfo}[2]{#2}
\providecommand{\eprint}[2][]{\url{#2}}

\bibitem{starbird2019disinformation}
\bibinfo{author}{Starbird, K.}, \bibinfo{author}{Arif, A.} \&
  \bibinfo{author}{Wilson, T.}
\newblock \bibinfo{journal}{\bibinfo{title}{Disinformation as collaborative
  work: Surfacing the participatory nature of strategic information
  operations}}.
\newblock {\emph{\JournalTitle{Proceedings of the ACM on Human-Computer
  Interaction}}} \textbf{\bibinfo{volume}{3}}, \bibinfo{pages}{1--26}
  (\bibinfo{year}{2019}).

\bibitem{pierri2020investigating}
\bibinfo{author}{Pierri, F.}, \bibinfo{author}{Artoni, A.} \&
  \bibinfo{author}{Ceri, S.}
\newblock \bibinfo{journal}{\bibinfo{title}{Investigating italian
  disinformation spreading on twitter in the context of 2019 european
  elections}}.
\newblock {\emph{\JournalTitle{PloS one}}} \textbf{\bibinfo{volume}{15}},
  \bibinfo{pages}{e0227821} (\bibinfo{year}{2020}).

\bibitem{soares2021hashtag}
\bibinfo{author}{Soares, F.~B.} \& \bibinfo{author}{Recuero, R.}
\newblock \bibinfo{journal}{\bibinfo{title}{Hashtag wars: political
  disinformation and discursive struggles on twitter conversations during the
  2018 brazilian presidential campaign}}.
\newblock {\emph{\JournalTitle{Social Media+ Society}}}
  \textbf{\bibinfo{volume}{7}}, \bibinfo{pages}{20563051211009073}
  (\bibinfo{year}{2021}).

\bibitem{xia2019disinformation}
\bibinfo{author}{Xia, Y.} \emph{et~al.}
\newblock \bibinfo{journal}{\bibinfo{title}{Disinformation, performed:
  Self-presentation of a russian ira account on twitter}}.
\newblock {\emph{\JournalTitle{Information, Communication \& Society}}}
  \textbf{\bibinfo{volume}{22}}, \bibinfo{pages}{1646--1664}
  (\bibinfo{year}{2019}).

\bibitem{casey1944goebbels}
\bibinfo{author}{Casey, R.~D.}
\newblock \bibinfo{title}{The goebbels experiment: A study of the nazi
  propaganda machine} (\bibinfo{year}{1944}).

\bibitem{berelson1947detecting}
\bibinfo{author}{Berelson, B.} \& \bibinfo{author}{De~Grazia, S.}
\newblock \bibinfo{journal}{\bibinfo{title}{Detecting collaboration in
  propaganda}}.
\newblock {\emph{\JournalTitle{Public Opinion Quarterly}}}
  \textbf{\bibinfo{volume}{11}}, \bibinfo{pages}{244--253}
  (\bibinfo{year}{1947}).

\bibitem{waddington2007anti}
\bibinfo{author}{Waddington, L.~L.}
\newblock \bibinfo{journal}{\bibinfo{title}{The anti-komintern and nazi
  anti-bolshevik propaganda in the 1930s}}.
\newblock {\emph{\JournalTitle{Journal of Contemporary History}}}
  \textbf{\bibinfo{volume}{42}}, \bibinfo{pages}{573--594}
  (\bibinfo{year}{2007}).

\bibitem{cabrera2020international}
\bibinfo{author}{Cabrera, M.~G.}
\newblock \bibinfo{journal}{\bibinfo{title}{International propaganda in spain
  during the first world war: State of the art and new contributions}}.
\newblock {\emph{\JournalTitle{Communication and the First World War}}}
  \bibinfo{pages}{188--218} (\bibinfo{year}{2020}).

\bibitem{zappavigna2014enacting}
\bibinfo{author}{Zappavigna, M.}
\newblock \bibinfo{journal}{\bibinfo{title}{Enacting identity in microblogging
  through ambient affiliation}}.
\newblock {\emph{\JournalTitle{Discourse \& Communication}}}
  \textbf{\bibinfo{volume}{8}}, \bibinfo{pages}{209--228}
  (\bibinfo{year}{2014}).

\bibitem{zhang2021assembling}
\bibinfo{author}{Zhang, Y.} \emph{et~al.}
\newblock \bibinfo{journal}{\bibinfo{title}{Assembling the networks and
  audiences of disinformation: How successful russian ira twitter accounts
  built their followings, 2015--2017}}.
\newblock {\emph{\JournalTitle{Journal of Communication}}}
  \textbf{\bibinfo{volume}{71}}, \bibinfo{pages}{305--331}
  (\bibinfo{year}{2021}).

\bibitem{garcia2018share}
\bibinfo{author}{Garc{\'\i}a-Perdomo, V.}, \bibinfo{author}{Salaverr{\'\i}a,
  R.}, \bibinfo{author}{Brown, D.~K.} \& \bibinfo{author}{Harlow, S.}
\newblock \bibinfo{journal}{\bibinfo{title}{To share or not to share: The
  influence of news values and topics on popular social media content in the
  united states, brazil, and argentina}}.
\newblock {\emph{\JournalTitle{Journalism studies}}}
  \textbf{\bibinfo{volume}{19}}, \bibinfo{pages}{1180--1201}
  (\bibinfo{year}{2018}).

\bibitem{posetti2018short}
\bibinfo{author}{Posetti, J.} \& \bibinfo{author}{Matthews, A.}
\newblock \bibinfo{journal}{\bibinfo{title}{A short guide to the history of
  ‘fake news’ and disinformation}}.
\newblock {\emph{\JournalTitle{International Center for Journalists}}}
  \textbf{\bibinfo{volume}{7}}, \bibinfo{pages}{2018--07}
  (\bibinfo{year}{2018}).

\bibitem{rajtmajer2020automated}
\bibinfo{author}{Rajtmajer, S.} \& \bibinfo{author}{Susser, D.}
\newblock \bibinfo{title}{Automated influence and the challenge of cognitive
  security}.
\newblock In \emph{\bibinfo{booktitle}{Proceedings of the 7th Symposium on Hot
  Topics in the Science of Security}}, \bibinfo{pages}{1--9}
  (\bibinfo{year}{2020}).

\bibitem{niblock2022understanding}
\bibinfo{author}{Niblock, I.}, \bibinfo{author}{Wallis, J.} \&
  \bibinfo{author}{Zhang, A.}
\newblock \bibinfo{title}{Understanding global disinformation and information
  operations} (\bibinfo{year}{2022}).

\bibitem{bradshaw2019global}
\bibinfo{author}{Bradshaw, S.} \& \bibinfo{author}{Howard, P.~N.}
\newblock \bibinfo{title}{The global disinformation order: 2019 global
  inventory of organised social media manipulation} (\bibinfo{year}{2019}).

\bibitem{bradshaw2018challenging}
\bibinfo{author}{Bradshaw, S.} \& \bibinfo{author}{Howard, P.~N.}
\newblock \bibinfo{journal}{\bibinfo{title}{Challenging truth and trust: A
  global inventory of organized social media manipulation}}.
\newblock {\emph{\JournalTitle{The computational propaganda project}}}
  \textbf{\bibinfo{volume}{1}}, \bibinfo{pages}{1--26} (\bibinfo{year}{2018}).

\bibitem{arnaudo2017computational}
\bibinfo{author}{Arnaudo, D.}
\newblock \bibinfo{title}{Computational propaganda in brazil: Social bots
  during elections} (\bibinfo{year}{2017}).

\bibitem{ong2018architects}
\bibinfo{author}{Ong, J.~C.} \& \bibinfo{author}{Caba{\~n}es, J. V.~A.}
\newblock \bibinfo{journal}{\bibinfo{title}{Architects of networked
  disinformation: Behind the scenes of troll accounts and fake news production
  in the philippines}}.
\newblock {\emph{\JournalTitle{Architects of networked disinformation: Behind
  the scenes of troll accounts and fake news production in the Philippines}}}
  (\bibinfo{year}{2018}).

\bibitem{vo2018rise}
\bibinfo{author}{Vo, N.} \& \bibinfo{author}{Lee, K.}
\newblock \bibinfo{title}{The rise of guardians: Fact-checking url
  recommendation to combat fake news}.
\newblock In \emph{\bibinfo{booktitle}{The 41st international ACM SIGIR
  conference on research \& development in information retrieval}},
  \bibinfo{pages}{275--284} (\bibinfo{year}{2018}).

\bibitem{sathe2020automated}
\bibinfo{author}{Sathe, A.}, \bibinfo{author}{Ather, S.}, \bibinfo{author}{Le,
  T.~M.}, \bibinfo{author}{Perry, N.} \& \bibinfo{author}{Park, J.}
\newblock \bibinfo{title}{Automated fact-checking of claims from wikipedia}.
\newblock In \emph{\bibinfo{booktitle}{Proceedings of the 12th Language
  Resources and Evaluation Conference}}, \bibinfo{pages}{6874--6882}
  (\bibinfo{year}{2020}).

\bibitem{volkova2019explaining}
\bibinfo{author}{Volkova, S.}, \bibinfo{author}{Ayton, E.},
  \bibinfo{author}{Arendt, D.~L.}, \bibinfo{author}{Huang, Z.} \&
  \bibinfo{author}{Hutchinson, B.}
\newblock \bibinfo{title}{Explaining multimodal deceptive news prediction
  models}.
\newblock In \emph{\bibinfo{booktitle}{Proceedings of the international AAAI
  conference on web and social media}}, vol.~\bibinfo{volume}{13},
  \bibinfo{pages}{659--662} (\bibinfo{year}{2019}).

\bibitem{garimella2020images}
\bibinfo{author}{Garimella, K.} \& \bibinfo{author}{Eckles, D.}
\newblock \bibinfo{journal}{\bibinfo{title}{Images and misinformation in
  political groups: Evidence from whatsapp in india}}.
\newblock {\emph{\JournalTitle{arXiv preprint arXiv:2005.09784}}}
  (\bibinfo{year}{2020}).

\bibitem{krishnamurthy2018deep}
\bibinfo{author}{Krishnamurthy, G.}, \bibinfo{author}{Majumder, N.},
  \bibinfo{author}{Poria, S.} \& \bibinfo{author}{Cambria, E.}
\newblock \bibinfo{journal}{\bibinfo{title}{A deep learning approach for
  multimodal deception detection}}.
\newblock {\emph{\JournalTitle{arXiv preprint arXiv:1803.00344}}}
  (\bibinfo{year}{2018}).

\bibitem{dawson2019russia}
\bibinfo{author}{Dawson, A.} \& \bibinfo{author}{Innes, M.}
\newblock \bibinfo{journal}{\bibinfo{title}{How russia's internet research
  agency built its disinformation campaign}}.
\newblock {\emph{\JournalTitle{The Political Quarterly}}}
  \textbf{\bibinfo{volume}{90}}, \bibinfo{pages}{245--256}
  (\bibinfo{year}{2019}).

\bibitem{myers2019china}
\bibinfo{author}{Myers, S.~L.} \& \bibinfo{author}{Mozur, P.}
\newblock \bibinfo{journal}{\bibinfo{title}{China is waging a disinformation
  war against hong kong protesters}}.
\newblock {\emph{\JournalTitle{The New York Times}}}
  \textbf{\bibinfo{volume}{13}} (\bibinfo{year}{2019}).

\bibitem{merhi2021information}
\bibinfo{author}{Merhi, M.}, \bibinfo{author}{Rajtmajer, S.} \&
  \bibinfo{author}{Lee, D.}
\newblock \bibinfo{journal}{\bibinfo{title}{Information operations in turkey:
  Manufacturing resilience with free twitter accounts}}.
\newblock {\emph{\JournalTitle{arXiv preprint arXiv:2110.08976}}}
  (\bibinfo{year}{2021}).

\bibitem{ferrara2017disinformation}
\bibinfo{author}{Ferrara, E.}
\newblock \bibinfo{journal}{\bibinfo{title}{Disinformation and social bot
  operations in the run up to the 2017 french presidential election}}.
\newblock {\emph{\JournalTitle{arXiv preprint arXiv:1707.00086}}}
  (\bibinfo{year}{2017}).

\bibitem{faris2017partisanship}
\bibinfo{author}{Faris, R.} \emph{et~al.}
\newblock \bibinfo{journal}{\bibinfo{title}{Partisanship, propaganda, and
  disinformation: Online media and the 2016 us presidential election}}.
\newblock {\emph{\JournalTitle{Berkman Klein Center Research Publication}}}
  \textbf{\bibinfo{volume}{6}} (\bibinfo{year}{2017}).

\bibitem{grinberg2019fake}
\bibinfo{author}{Grinberg, N.}, \bibinfo{author}{Joseph, K.},
  \bibinfo{author}{Friedland, L.}, \bibinfo{author}{Swire-Thompson, B.} \&
  \bibinfo{author}{Lazer, D.}
\newblock \bibinfo{journal}{\bibinfo{title}{Fake news on twitter during the
  2016 us presidential election}}.
\newblock {\emph{\JournalTitle{Science}}} \textbf{\bibinfo{volume}{363}},
  \bibinfo{pages}{374--378} (\bibinfo{year}{2019}).

\bibitem{hutchins1995cognition}
\bibinfo{author}{Hutchins, E.}
\newblock \bibinfo{journal}{\bibinfo{title}{Cognition in the wild mit press}}.
\newblock {\emph{\JournalTitle{Cambridge, MA}}} \textbf{\bibinfo{volume}{15}}
  (\bibinfo{year}{1995}).

\bibitem{sarcevic2012beacons}
\bibinfo{author}{Sarcevic, A.} \emph{et~al.}
\newblock \bibinfo{title}{" beacons of hope" in decentralized coordination:
  learning from on-the-ground medical twitterers during the 2010 haiti
  earthquake}.
\newblock In \emph{\bibinfo{booktitle}{Proceedings of the ACM 2012 conference
  on computer supported cooperative work}}, \bibinfo{pages}{47--56}
  (\bibinfo{year}{2012}).

\bibitem{schoch2022coordination}
\bibinfo{author}{Schoch, D.}, \bibinfo{author}{Keller, F.~B.},
  \bibinfo{author}{Stier, S.} \& \bibinfo{author}{Yang, J.}
\newblock \bibinfo{journal}{\bibinfo{title}{Coordination patterns reveal online
  political astroturfing across the world}}.
\newblock {\emph{\JournalTitle{Scientific reports}}}
  \textbf{\bibinfo{volume}{12}}, \bibinfo{pages}{1--10} (\bibinfo{year}{2022}).

\bibitem{del2016spreading}
\bibinfo{author}{Del~Vicario, M.} \emph{et~al.}
\newblock \bibinfo{journal}{\bibinfo{title}{The spreading of misinformation
  online}}.
\newblock {\emph{\JournalTitle{Proceedings of the national academy of
  Sciences}}} \textbf{\bibinfo{volume}{113}}, \bibinfo{pages}{554--559}
  (\bibinfo{year}{2016}).

\bibitem{wang2020fake}
\bibinfo{author}{Wang, R.}, \bibinfo{author}{He, Y.}, \bibinfo{author}{Xu, J.}
  \& \bibinfo{author}{Zhang, H.}
\newblock \bibinfo{journal}{\bibinfo{title}{Fake news or bad news? toward an
  emotion-driven cognitive dissonance model of misinformation diffusion}}.
\newblock {\emph{\JournalTitle{Asian Journal of Communication}}}
  \textbf{\bibinfo{volume}{30}}, \bibinfo{pages}{317--342}
  (\bibinfo{year}{2020}).

\bibitem{chen2015deterring}
\bibinfo{author}{Chen, X.}, \bibinfo{author}{Sin, S.-C.~J.},
  \bibinfo{author}{Theng, Y.-L.} \& \bibinfo{author}{Lee, C.~S.}
\newblock \bibinfo{journal}{\bibinfo{title}{Deterring the spread of
  misinformation on social network sites: A social cognitive theory-guided
  intervention}}.
\newblock {\emph{\JournalTitle{Proceedings of the Association for Information
  Science and Technology}}} \textbf{\bibinfo{volume}{52}},
  \bibinfo{pages}{1--4} (\bibinfo{year}{2015}).

\bibitem{pennycook2020fighting}
\bibinfo{author}{Pennycook, G.}, \bibinfo{author}{McPhetres, J.},
  \bibinfo{author}{Zhang, Y.}, \bibinfo{author}{Lu, J.~G.} \&
  \bibinfo{author}{Rand, D.~G.}
\newblock \bibinfo{journal}{\bibinfo{title}{Fighting covid-19 misinformation on
  social media: Experimental evidence for a scalable accuracy-nudge
  intervention}}.
\newblock {\emph{\JournalTitle{Psychological science}}}
  \textbf{\bibinfo{volume}{31}}, \bibinfo{pages}{770--780}
  (\bibinfo{year}{2020}).

\bibitem{pennycook2021psychology}
\bibinfo{author}{Pennycook, G.} \& \bibinfo{author}{Rand, D.~G.}
\newblock \bibinfo{journal}{\bibinfo{title}{The psychology of fake news}}.
\newblock {\emph{\JournalTitle{Trends in cognitive sciences}}}
  \textbf{\bibinfo{volume}{25}}, \bibinfo{pages}{388--402}
  (\bibinfo{year}{2021}).

\bibitem{berger2005idea}
\bibinfo{author}{Berger, J.~A.} \& \bibinfo{author}{Heath, C.}
\newblock \bibinfo{journal}{\bibinfo{title}{Idea habitats: How the prevalence
  of environmental cues influences the success of ideas}}.
\newblock {\emph{\JournalTitle{Cognitive Science}}}
  \textbf{\bibinfo{volume}{29}}, \bibinfo{pages}{195--221}
  (\bibinfo{year}{2005}).

\bibitem{penney2017citizen}
\bibinfo{author}{Penney, J.}
\newblock \emph{\bibinfo{title}{The citizen marketer: Promoting political
  opinion in the social media age}} (\bibinfo{publisher}{Oxford University
  Press}, \bibinfo{year}{2017}).

\bibitem{jakubowski2019s}
\bibinfo{author}{Jakubowski, G.}
\newblock \bibinfo{journal}{\bibinfo{title}{What’s not to like? social media
  as information operations force multiplier}}.
\newblock {\emph{\JournalTitle{Joint Force Quarterly}}}
  \textbf{\bibinfo{volume}{3}}, \bibinfo{pages}{8--17} (\bibinfo{year}{2019}).

\bibitem{lee2014discovering}
\bibinfo{author}{Lee, A.~J.}, \bibinfo{author}{Yang, F.-C.},
  \bibinfo{author}{Tsai, H.-C.} \& \bibinfo{author}{Lai, Y.-Y.}
\newblock \bibinfo{journal}{\bibinfo{title}{Discovering content-based
  behavioral roles in social networks}}.
\newblock {\emph{\JournalTitle{Decision Support Systems}}}
  \textbf{\bibinfo{volume}{59}}, \bibinfo{pages}{250--261}
  (\bibinfo{year}{2014}).

\bibitem{han2014balanced}
\bibinfo{author}{Han, S.}, \bibinfo{author}{Zhuang, F.}, \bibinfo{author}{He,
  Q.} \& \bibinfo{author}{Shi, Z.}
\newblock \bibinfo{title}{Balanced seed selection for budgeted influence
  maximization in social networks}.
\newblock In \emph{\bibinfo{booktitle}{Pacific-Asia Conference on Knowledge
  Discovery and Data Mining}}, \bibinfo{pages}{65--77}
  (\bibinfo{organization}{Springer}, \bibinfo{year}{2014}).

\bibitem{lorrain1971structural}
\bibinfo{author}{Lorrain, F.} \& \bibinfo{author}{White, H.~C.}
\newblock \bibinfo{journal}{\bibinfo{title}{Structural equivalence of
  individuals in social networks}}.
\newblock {\emph{\JournalTitle{The Journal of mathematical sociology}}}
  \textbf{\bibinfo{volume}{1}}, \bibinfo{pages}{49--80} (\bibinfo{year}{1971}).

\bibitem{forestier2012roles}
\bibinfo{author}{Forestier, M.}, \bibinfo{author}{Stavrianou, A.},
  \bibinfo{author}{Velcin, J.} \& \bibinfo{author}{Zighed, D.~A.}
\newblock \bibinfo{journal}{\bibinfo{title}{Roles in social networks:
  Methodologies and research issues}}.
\newblock {\emph{\JournalTitle{Web Intelligence and Agent Systems: An
  international Journal}}} \textbf{\bibinfo{volume}{10}},
  \bibinfo{pages}{117--133} (\bibinfo{year}{2012}).

\bibitem{rundin2021multifaceted}
\bibinfo{author}{Rundin, K.} \& \bibinfo{author}{Colliander, J.}
\newblock \bibinfo{journal}{\bibinfo{title}{Multifaceted influencers: toward a
  new typology for influencer roles in advertising}}.
\newblock {\emph{\JournalTitle{Journal of Advertising}}}
  \textbf{\bibinfo{volume}{50}}, \bibinfo{pages}{548--564}
  (\bibinfo{year}{2021}).

\bibitem{TwitterIO}
\bibinfo{title}{Twitter moderation research consortium}.
\newblock
  \bibinfo{howpublished}{\url{https://transparency.twitter.com/en/reports/moderation-research.html}}.
\newblock \bibinfo{note}{Accessed: 2023-03-20}.

\bibitem{TwitterBlog}
\bibinfo{title}{Disclosing new data to our archive of information operations}.
\newblock
  \bibinfo{howpublished}{\url{https://blog.twitter.com/en_us/topics/company/2019/info-ops-disclosure-data-september-2019}}.
\newblock \bibinfo{note}{Accessed: 2023-03-20}.

\bibitem{squartini2013reciprocity}
\bibinfo{author}{Squartini, T.}, \bibinfo{author}{Picciolo, F.},
  \bibinfo{author}{Ruzzenenti, F.} \& \bibinfo{author}{Garlaschelli, D.}
\newblock \bibinfo{journal}{\bibinfo{title}{Reciprocity of weighted networks}}.
\newblock {\emph{\JournalTitle{Scientific reports}}}
  \textbf{\bibinfo{volume}{3}}, \bibinfo{pages}{2729} (\bibinfo{year}{2013}).

\bibitem{zappavigna2011ambient}
\bibinfo{author}{Zappavigna, M.}
\newblock \bibinfo{journal}{\bibinfo{title}{Ambient affiliation: A linguistic
  perspective on twitter}}.
\newblock {\emph{\JournalTitle{New media \& society}}}
  \textbf{\bibinfo{volume}{13}}, \bibinfo{pages}{788--806}
  (\bibinfo{year}{2011}).

\bibitem{erz2018hashtags}
\bibinfo{author}{Erz, A.}, \bibinfo{author}{Marder, B.} \&
  \bibinfo{author}{Osadchaya, E.}
\newblock \bibinfo{journal}{\bibinfo{title}{Hashtags: Motivational drivers,
  their use, and differences between influencers and followers}}.
\newblock {\emph{\JournalTitle{Computers in Human Behavior}}}
  \textbf{\bibinfo{volume}{89}}, \bibinfo{pages}{48--60}
  (\bibinfo{year}{2018}).

\bibitem{conway2013twitter}
\bibinfo{author}{Conway, B.~A.}, \bibinfo{author}{Kenski, K.} \&
  \bibinfo{author}{Wang, D.}
\newblock \bibinfo{journal}{\bibinfo{title}{Twitter use by presidential primary
  candidates during the 2012 campaign}}.
\newblock {\emph{\JournalTitle{American Behavioral Scientist}}}
  \textbf{\bibinfo{volume}{57}}, \bibinfo{pages}{1596--1610}
  (\bibinfo{year}{2013}).

\bibitem{martin2016hashtags}
\bibinfo{author}{Mart{\'\i}n, E.~G.}, \bibinfo{author}{Lavesson, N.} \&
  \bibinfo{author}{Doroud, M.}
\newblock \bibinfo{journal}{\bibinfo{title}{Hashtags and followers: An
  experimental study of the online social network twitter}}.
\newblock {\emph{\JournalTitle{Social Network Analysis and Mining}}}
  \textbf{\bibinfo{volume}{6}}, \bibinfo{pages}{1--15} (\bibinfo{year}{2016}).

\bibitem{benkler2018network}
\bibinfo{author}{Benkler, Y.}, \bibinfo{author}{Faris, R.} \&
  \bibinfo{author}{Roberts, H.}
\newblock \emph{\bibinfo{title}{Network propaganda: Manipulation,
  disinformation, and radicalization in American politics}}
  (\bibinfo{publisher}{Oxford University Press}, \bibinfo{year}{2018}).

\bibitem{berghel2017oh}
\bibinfo{author}{Berghel, H.}
\newblock \bibinfo{journal}{\bibinfo{title}{Oh, what a tangled web: Russian
  hacking, fake news, and the 2016 us presidential election}}.
\newblock {\emph{\JournalTitle{Computer}}} \textbf{\bibinfo{volume}{50}},
  \bibinfo{pages}{87--91} (\bibinfo{year}{2017}).

\end{thebibliography}

\section*{Data availability statement}

The datasets described herein are readily available and publicly accessible through Twitter's Moderation Research Consortium at the following URL: \url{https://transparency.twitter.com/en/reports/moderation-research.html}. Datasets can be downloaded instantly upon providing a valid email address.

\section*{Acknowledgements}
The authors received no specific funding for this work. 

\section*{Author contributions}

X.W. conceived and conducted experiments, analyzed results, and drafted the paper,  J.L. conducted experiments, analyzed results, and contributed to writing the paper, E.S. conducted experiments and analyzed results, S.R. conceived research questions and experiments, and contributed to writing the paper. All authors reviewed the manuscript. 

\section*{Competing interests}
The authors declare no competing interests.

\end{document}